\documentclass[11pt]{article}
\usepackage[english]{babel}
\usepackage{graphicx}
\usepackage{times}
\usepackage{url}
\usepackage{amsmath}
\usepackage{amsfonts}
\usepackage{amssymb}

\begin{document}

\date{}
\title{Dynamical systems analysis of stack filters}
\author{Matti Nykter$^{1,2}$, Juha Kesseli$^{3,2}$, Ilya Shmulevich$^{1}$ \\
%EndAName
{\normalsize {$^{1}$Institute for Systems Biology, Seattle, WA, 98103, USA}}%
\\
{\normalsize {$^{2}$Department of Signal Processing, Tampere University of
Technology},}\\
{\normalsize {33720 Tampere, Finland}}\\
{\normalsize {$^{3}$Group for Neural Theory, D{\'{e}}partement d'Etudes
Cognitives},}\\
{\normalsize {Ecole Normale Sup{\'e}rieure, 75005 Paris, France}}}
\maketitle

\begin{abstract}
We study classes of dynamical systems that can be obtained by constructing
recursive networks with monotone Boolean functions. Stack filters in
nonlinear signal processing are special cases of such systems. We show an
analytical connection between coefficients used to optimize the statistical
properties of stack filters and their sensitivity, a measure that can be
used to characterize the dynamical properties of Boolean networks
constructed from the corresponding monotone functions. A connection is made
between the rank selection probabilities (RSPs) and the sensitivity. We also
examine the dynamical behavior of monotone functions corresponding to
filters that are optimal in terms of their noise suppression capability, and
find that the optimal filters are dynamically chaotic. This contrasts with
the optimal information preservation properties of critical networks in the
case of small perturbations in Boolean networks, and highlights the
difference between such perturbations and those corresponding to noise. We
also consider a generalization of Boolean networks that is obtained by
utilizing stack filters on continuous-valued states. It can be seen that for
such networks, the dynamical regime can be changed when binary variables are
made continuous.
\end{abstract}

\thispagestyle{empty}

\section{Introduction}

Biological systems can be viewed as dynamical systems of biomolecular
interactions that process information from their environment and mount
diverse, yet specific responses, striking a balance between stability and
adaptability. Living systems need to remain stable under variable
environmental conditions and yet be able to evolve and respond to specific
stimuli. This trade-off between adaptability and stability can be captured,
in the context of dynamical systems analysis, by an order parameter that
quantifies the sensitivity of the system to perturbations, such as
environmental and molecular noise.

Recent experimental evidence has been mounting in support of a long-standing
hypothesis stating that living systems operate at the critical regime
between ordered and disordered behavior \cite{kauffman93, serra04, ramo06,
nykter08b}. Furthermore, critical systems have been shown to be optimal in
several ways. For example, they are able to store a maximal amount of
information \cite{krawitz07} and propagate the information with minimal
information loss in a noisy environment \cite{ribeiro08}. Additionally,
critical systems exhibit the most complex relationships between their
structure and dynamics \cite{nykter08}. Thus, optimality of a dynamical
system can be considered as a trade-off between noise suppression and detail
preservation. Such a trade-off is also the central goal of designing digital
filters in signal processing. Indeed, digital filters can be studied in the
context of dynamical systems theory \cite{shmulevich03c}.

In this contribution, we consider the well known class of nonlinear digital
filters, called stack filters. There exists a well established statistical
theory for the optimization of stack filters for a given noise distribution 
\cite{astola97, gabbouj92}. Statistical properties of stack filters can be
represented by a set of coefficients and these can be used to design an
optimal class of stack filters that minimizes some statistical criterion,
such as noise variance at the output of the filter \cite{gabbouj90,
gabbouj91}. These coefficients can also be used to derive statistics that
quantify the properties of the filters. Since the output of a stack filter
is always one of the inputs, rank selection probabilities (RSP) quantify
robustness and the noise suppression of the filter, specifying the
probability that a sample of a particular rank in the input window will be
used as the output of the filter; while sample selection probabilities (SSP)
can be used to study the filter's detail preservation, specifying the
probability that a particular sample in the input window appears at the
output \cite{shmulevich00}. Thus, these statistics capture the trade-off
between robustness and detail preservation. For example, the median filter
is extremely robust, but preserves details rather poorly, since its RSPs are
all zero, except the center one, which is equal to one (the one
corresponding to the median), while the SSPs are all equal (i.e., the
uniform distribution). At the other extreme, the identity filter has the
highest level of detail preservation, but very poor robustness to noise,
since the RSPs are all equal, while the SSPs are all zero, except for the
center one, which is equal to one (the one corresponding to the center
position of the window).

We will show an analytical relationship between stack filter coefficients,
rank selection properties, and the order parameter (essentially, Lyapunov
exponent) that characterizes the behavior of the corresponding dynamical
system, thus connecting the statistical optimization of stack filters with
dynamical systems theory.

\section{Background}

\subsection{Boolean networks and dynamical regimes}

A Boolean network model is a conceptually simple dynamical system model
where each node can be in only two possible states, on or off. Despite the
apparent simplicity, this model class is able to produce highly complex
behavior, for example, in the form of a phase transition between two
dynamical regimes in which small perturbations are either attenuated or
amplified. A Boolean network model can be defined as follows.

Let $s_{i}(t)\in \{0,1\}$, $i=1,\ldots ,N$, where $N$ is the number of nodes
in the network, be the state of $i$th node in the Boolean network at time $t$%
. The state of this node at time $t+1$ is determined by the states of nodes $%
j_{1},j_{2},{\ldots },j_{k_{i}}$ at time $t$ as%
\begin{equation}
s_{i}(t+1)=f_{i}(s_{j_{1}}(t),s_{j_{2}}(t),{\ldots },s_{j_{k_{i}}}(t)),
\label{eq:bnupdate}
\end{equation}%
where $f_{i}:\{0,1\}^{k_{i}}\rightarrow \{0,1\}$ is a Boolean function of $%
k_{i}$ variables. A binary vector $s\left( t\right) =\left( s_{1}\left(
t\right) ,\ldots ,s_{N}\left( t\right) \right) $ is said to be the state of
the network at time $t.$ In the classical model, all nodes are updated
synchronously as the system transitions from state $s\left( t\right) $ to
state $s\left( t+1\right) $ \cite{kauffman93}. The state transitions are
determined by the multi-output Boolean function $F=(f_{1},{\ldots },f_{N})$
as $s(t+1)=F(s(t)),$ where $f_{i}$ is the Boolean function of node $i$ with
predetermined connections from the nodes $j_{1},j_{2},{\ldots },j_{k_{i}}.$
It should be noted that this model can directly be generalized to a larger
alphabet by defining $s_{i}(t)\in \{0,\ldots ,L-1\}$ and $f_{i}:\{0,\ldots
,L-1\}^{k_{i}}\rightarrow \{0,\ldots ,L-1\}$, where $L$ is the size of the
alphabet. We restrict our attention to the case $L=2.$

To construct a Boolean network, the inputs $j_{1},j_{2},{\ldots },j_{k_{i}}$
for each node $i$ needs to be determined. This can be done by selecting the
inputs randomly among all $N$ nodes or by selecting the inputs using some
systematic pattern. The number of inputs $k_{i}$ can be endowed with a
probability distribution, such as the power-law \cite{barabasi99, aldana03b}
or Poisson distribution, with a mean $K=E\left[ k_{i}\right] ,$ known as the
average connectivity of the network.

Once the connections have been set, we can choose a Boolean function $f_{i}$
for each node. Functions can be parameterized by the bias $b=E\left[ f_{i}%
\right] $, the probability that the function outputs one on an arbitrary
input vector. If \mbox{$b=0.5$}, then the function is said to be unbiased.
The functions can be selected randomly among all $2^{2^{k_{i}}}$ Boolean
functions or they can be selected from some class of functions \cite%
{stauffer87,shmulevich03, kauffman00, harris02}. If both the functions and
connections are selected randomly, then the obtained network is called a
random Boolean network (RBN) \cite{kauffman93}.

Since a Boolean network is a discrete system, it has a finite state space.
Thus, every state trajectory, that is, a path through the state space from
any initial state, will eventually return to one of the previously visited
states. This kind of a state cycle where the same states are repeated
infinitely is known as an attractor cycle and the states within the cycle
are called attractor states. A set of states that leads to the same
attractor is called the basin of attraction \cite{wuensche99}.

Boolean networks, as models of dynamical systems, can operate in the ordered
or chaotic regimes, or at the phase transition boundary between these two
regimes \cite{aldana03}. This phase transition regime has been referred to
as the edge of chaos \cite{kauffman93}. When a network is operating in the
ordered regime, it is intrinsically robust while its dynamical behavior is
simple. The robustness can be observed through both the structural and
transient perturbations. Perturbations have a small effect on the behavior
of the network. Networks in the chaotic regime, on the other hand, are
extremely sensitive to perturbations. Even a small perturbation will quickly
propagate through the entire network. Thus, networks in the chaotic regime
are not robust in that they are not able to coordinate macroscopic behavior
under perturbations. A phase transition between the ordered and chaotic
regimes represents a trade-off between the need for stability and the need
to have a wide range of dynamical behavior to respond to variable
perturbations \cite{kauffman93}.

By varying the parameters $K$ and $b$ in the random Boolean network model, a
dynamical phase transition can take place. The average sensitivity of the
Boolean network 
\begin{equation}
\sigma =2b(1-b)K  \label{eq:sensitivity}
\end{equation}%
can be used to determine the dynamical regime. If $\sigma >1$ then the
system is chaotic and for $\sigma <1$ the system is ordered \cite{derrida86,
luque97, luque00, shmulevich04}. It is easy to see that for unbiased random
Boolean networks, the critical connectivity is $K_{c}=2$.

In a sense, critical systems are maximally responsive to the useful
information in their environment while being able to reliably execute their
behaviors in the presence of uninformative variation in this environment. A
hallmark of critical behavior is the spontaneous emergence of complex and
coordinated macroscopic behavior in the form of long-range spatial or
temporal correlations. Such coordination across many scales enables
information to propagate over time from one part of the system to another
with a high degree of specificity and sensitivity. These aspects of
criticality support the idea that living cells, as complex dynamical systems
of interacting biomolecules, are dynamically critical systems.

\subsection{Stack Filters}

Stack filters constitute an important class of nonlinear filters based on
monotone Boolean functions \cite{wendt86}. Statistical properties of stack
filters have been studied in terms of output distributions and moments for
independent and identically distributed input signals \cite{agaian95,
gabbouj90}. Consequently, it becomes possible to optimize stack filters in
the mean square sense. In other words, the knowledge of the input
distribution allows one to find a stack filter or a set of stack filters
that minimize the output variance.

Rank selection probabilities (RSP) and sample selection probabilities (SSP)
are probabilities that the output equals a sample with a certain rank and
certain time-index in the filter window, respectively. The output
distribution of a stack filter can be expressed in terms of its RSP's. On
the other hand, SSP's give us information about the temporal behavior of
stack filters. This information is important for examining the detail
preservation properties of stack filters.

Let $\alpha =\left( \alpha _{1},\cdots ,\alpha _{n}\right) $ and $\beta
=\left( \beta _{1},\cdots ,\beta _{n}\right) $ be two different $n$-element
binary vectors. We say that $\alpha $ precedes $\beta $, denoted as $\alpha
\prec \beta $, if $\alpha _{i}\leq \beta _{i}$ for every $i$, $1\leq i\leq n$%
. If $\alpha \nprec \beta $ and $\beta \nprec \alpha $, then $\alpha $ and $%
\beta $ are said to be incomparable. Relative to the predicate $\prec $, the
set of all binary vectors of a given length is a partially ordered set. A
Boolean function $f\left( x_{1},\cdots ,x_{n}\right) $ is called monotone if
for any two vectors $\alpha $ and $\beta $ such that $\alpha \prec \beta $,
we have $f\left( \alpha \right) \leq f(\beta )$. The class of monotone
Boolean functions is one of the most widely used and studied classes of
Boolean functions.

Let $E^{n}$ denote the Boolean $n$-cube, that is, a graph with $2^{n}$
vertices each of which is labeled by an $n$-element binary vector. Two
vertices $\alpha =\left( \alpha _{1},\cdots ,\alpha _{n}\right) $ and $\beta
=\left( \beta _{1},\cdots ,\beta _{n}\right) $ are connected by an edge if
and only if the Hamming distance $\rho (\alpha ,\beta )=\sum_{i=1}^{n}\left(
\alpha _{i}\oplus \beta _{i}\right) =1,$ where $\oplus $ is addition modulo
2 (exclusive OR). The set of those vectors from $E^{n}$ in which there are
exactly $k$ units, $0\leq k\leq n$, is called the $k$th layer of $E^{n}$ and
is denoted by $E^{n,k}$. The Hamming weight of the vector $x\in \{0,1\}^{n}$
is $w(x)=\rho (x,0),$ where $0$ is the all-zero vector.

A stack filter is defined by a monotone Boolean function. A continuous stack
filter $S_{f}(\cdot )$, based on function $f$, is obtained by replacing
conjunction $(\cdot )$ and disjunction $(+)$ operations by $\min (\cdot )$
and $\max (\cdot )$ operations, respectively. For example, the monotone
Boolean function $f(x_{1},x_{2},x_{3})=x_{2}+x_{1}x_{3}$ corresponds to the
continuous stack filter $S_{f}(X_{1},X_{2},X_{3})=\max (X_{2},\min
(X_{1},X_{3})).$

Suppose that the input variables of some stack filter $S_{f}(\cdot )$ are
i.i.d. random variables with distribution $F(t)=\Pr \{X_{i}\leq t\}$. Then,
it is well known \cite{agaian95} that the output $Y=S_{f}(X_{1},{\ldots }%
,X_{n})$ of the stack filter has output distribution 
\begin{equation}
\Phi (t)=\sum_{i=0}^{n-1}A_{i}(1-F(t))^{i}F(t)^{n-i}
\end{equation}%
where 
\begin{equation}
A_{i}=|\{x\in E^{n,i}:f(x)=0\}|.  \label{eqAi}
\end{equation}

It is known \cite{kuosmanen94, shmulevich02b} that rank selection
probabilities $p_{i}=\Pr \{Y=X_{(i)}\},$ where $X_{(i)}$ is the $i$th order
statistic, are obtained as, 
\begin{equation}
p_{i}=\frac{A_{n-i}}{{\binom{n}{n-i}}}-\frac{A_{n-i+1}}{{\binom{n}{n-i+1}}}.
\label{eq:rank}
\end{equation}%
It can also be shown that the output distribution function of the filter can
be expressed in terms of the RSPs as 
\begin{equation}
\Phi (t)=\sum_{i=1}^{n}p_{i}F_{(i)}(t),
\end{equation}%
where $F_{(i)}(t)$ is the cumulative distribution function of the $i$%
th-order statistic for i.i.d. inputs \cite{astola97}.

Let $s_{j}=\Pr \{Y=X_{j}\}$ be the $j$th SSP and $s=(s_{1},{\ldots },s_{n})$
be the vector of SSPs. In \cite{egiazarian96}, it was shown that $%
s_{j}=d_{j}(1)-d_{j}(0)$, where 
\begin{equation}
d_{j}(k)=\sum_{x\in f^{-1}(1)|x_{j}=k}\left[ n{\binom{{n-1}}{{w(x)-k}}}%
\right] ^{-1}
\end{equation}%
and $f^{-1}(k)=\{x\in E^{n}:f(x)=k\}$.

For a given input distribution, the mean square optimization of stack
filters can be performed as follows. The variance $\mu
_{2}=E\{(Y-E\{Y\})^{2}\}$ of the output $Y$ of the stack filter can be
written as 
\begin{equation}
\mu _{2}=\sum_{i=0}^{n-1}A_{i}M(F,2,n,i)-\left(
\sum_{i=0}^{n-1}A_{i}M(F,1,n,i)\right) ^{2},  \label{eq:obj}
\end{equation}%
where 
\begin{equation}
M(F,k,n,i)=\int_{-\infty }^{\infty }x^{k}\frac{d}{dt}%
((1-F(t))^{i}F(t)^{n-i})dx.
\end{equation}

Thus, for a given noise distribution, the goal of optimization is to find
parameters $A_{i}$ (equivalently, the rank selection probabilities) such
that the objective function (\ref{eq:obj}) is minimized. An optimal stack
filter, specified by monotone Boolean function $f$, is then constructed
using the obtained parameters $A_{i}$. As a result of optimization, we
obtain a set of parameters $A_{i}$ that define a class of stack filters. In
a given class, all stack filters are statistically equivalent since they all
possess the same parameters $A_{i}$. Thus, they will also have the same rank
selection probabilities $p_{i}$. An approach to select the best filter, in
terms of its ability to preserve details, from the class of such
statistically equivalent optimal stack filters has been proposed in \cite%
{shmulevich00}.

\subsection{Average sensitivity of Boolean functions}

Let $f:\left\{ 0,1\right\} ^{n}\rightarrow \left\{ 0,1\right\} $ be a
Boolean function of $n$ variables $x_{1},\ldots ,x_{n}$. Let 
\begin{equation}
\partial f\left( x\right) /\partial x_{j}=f\left( x^{\left( j,0\right)
}\right) \oplus f\left( x^{\left( j,1\right) }\right)
\end{equation}%
be the partial derivative of $f$ with respect to $x_{j}$, where $x^{\left(
j,k\right) }=\left( x_{1},\ldots ,x_{j-1},k,x_{j+1},\ldots x_{n}\right) $, $%
\,k=0,1$. Clearly, the partial derivative is a Boolean function itself that
specifies whether a change in the $j$th input causes a change in the
original function $f$. Now, the activity of variable $x_{j}$ in function $f$
can be defined as 
\begin{equation}
\alpha _{j}^{f}=\frac{1}{2^{n}}\sum_{x\in \left\{ 0,1\right\} ^{n}}\partial
f\left( x\right) /\partial x_{j}.  \label{eqactivity}
\end{equation}%
Note that although the vector $x$ consists of $n$ components (variables),
the $j$th variable is fictitious in $\partial f\left( x\right) /\partial
x_{j}$. A variable $x_{j}$ is fictitious in $f$ if $f\left( x^{\left(
j,0\right) }\right) =f\left( x^{\left( j,1\right) }\right) $ for all $%
x^{\left( j,0\right) }$ and $x^{\left( j,1\right) }$. For an $n$-variable
Boolean function $f$, we can form its activity vector $\alpha ^{f}=\left[
\alpha _{1}^{f},\ldots ,\alpha _{n}^{f}\right] $. It is easy to see that $%
0\leq \alpha _{j}^{f}\leq 1,$ for any $j=1,\ldots ,n$. In fact, we can
consider $\alpha _{j}^{f}$ to be a probability that toggling the $j$th input
bit changes the function value, when the input vectors $x$ are distributed
uniformly over $\left\{ 0,1\right\} ^{n}$. Since we're in the binary
setting, the activity is also the expectation of the partial derivative with
respect to the uniform distribution: $\alpha _{j}^{f}=E\left[ \partial
f\left( x\right) /\partial x_{j}\right] $.

Another important quantity is the sensitivity of a Boolean function $f$,
which measures how sensitive the output of the function is to changes in the
inputs. The sensitivity $s^{f}\left( x\right) $ of $f$ on vector $x$ is
defined as the number of Hamming neighbors of $x$ on which the function
value is different than on $x$ (two vectors are Hamming neighbors if they
differ in only one component). That is, 
\begin{align}
s^{f}\left( x\right) & =\left\vert \left\{ i\in \left\{ 1,\ldots ,n\right\}
:f\left( x\oplus e_{i}\right) \neq f\left( x\right) \right\} \right\vert \\
& =\sum_{i=1}^{n}\chi \left[ f\left( x\oplus e_{i}\right) \neq f\left(
x\right) \right] ,  \notag
\end{align}%
where $e_{i}$ is the unit vector with 1 in the $i$th position and 0s
everywhere else and $\chi \left[ A\right] $ is an indicator function that is
equal to 1 if and only if $A$ is true. The \textit{average sensitivity} $%
s^{f}$ is defined by taking the expectation of $s^{f}\left( x\right) $ with
respect to the distribution of $x$. It is easy to see that under the uniform
distribution, the average sensitivity is equal to the sum of the activities: 
\begin{align}
s^{f}& =E\left[ s^{f}\left( x\right) \right] =\sum_{i=1}^{n}E\left[ \chi %
\left[ f\left( x\oplus e_{i}\right) \neq f\left( x\right) \right] \right]
\label{eqavgsens} \\
& =\sum_{i=1}^{n}\alpha _{i}^{f}.  \notag
\end{align}%
Therefore, $s^{f}$ is a number between $0$ and $n$. For a random Boolean
network, the mean of the average sensitivities $s^{f_{i}}$ of the Boolean
functions $f_{i}$, $i=1,\ldots ,N,$ corresponding to each of the $N$ nodes
is the network average sensitivity $\sigma $ given in (\ref{eq:sensitivity}) 
\cite{shmulevich04}.

\section{Average sensitivity and rank selection probabilities}

As defined in (\ref{eqavgsens}), to compute the sensitivity of a monotone
function $f$ with $n$ variables, we need to go through all the one-bit
changes (from each input vector $x\in E^{n}$) and see how many times, doing
this, the output changes. We can divide the task into $n$ sub-tasks, in each
of which we only consider bit changes (perturbations) from a vector $x$ with 
$w(x)=k$ into a vector $y$ with $w(y)=k+1$ ($k=0,1,{\ldots },n-1$). We can
denote this set of perturbations $(x,y)$ by $F_{k}$ for simplicity.

For each of the ${\binom{n}{k}}$ vectors $x\in E^{n,k},$ it holds that 
\begin{equation*}
|{z:(x,z)\in F_{k}|}=n-k,
\end{equation*}%
and conversely, for each vector $y$ with $w(y)=k+1$ it holds that 
\begin{equation*}
|{z:(z,y)\in F_{k}|}=k+1.
\end{equation*}%
In total, 
\begin{equation}
\left\vert F_{k}\right\vert ={\binom{n}{k}}(n-k)={\binom{n}{k+1}}(k+1),
\label{eqtotal}
\end{equation}%
so that there are this many perturbations to check in this sub-task. Since
we know the total number $\sum_{k}F_{k}=n\times 2^{n-1},$ we can just as
well look for bit changes in each $F_{k}$ that do not change the output of $%
f $.

Based on the monotonicity of $f$, we know that for each $x\in E^{n,k}$ and $%
f(x)=1,$ all perturbations $(x,y)$ in $F_{k}$ will not alter the output of
the function (since $f(y)=1$ as well). There are ${\binom{n}{k}}-A_{k}$ such
states and hence 
\begin{equation}
\left[ {\binom{n}{k}}-A_{k}\right](n-k)  \label{eq1}
\end{equation}
such non-output-altering perturbations in $F_{k}$.

On the other hand, we know that for each $y\in E^{n,k+1}$ and $f(y)=0,$ all
perturbations $(x,y)$ will also be non-output-altering (since $f(x)=0$ due
to monotonicity). There are $A_{k+1}$ such states and hence 
\begin{equation}
A_{k+1}(k+1)  \label{eq2}
\end{equation}
such perturbations in $F_{k}$.

The number of output-altering perturbations in $F_{k}$ can now be computed
by taking out the contributions of (\ref{eq1}) and (\ref{eq2}) from the
total number in (\ref{eqtotal}), i.e. 
\begin{equation*}
{\binom{n}{k+1}}(k+1)-\left[ {\binom{n}{k}}-A_{k}\right] (n-k)-A_{k+1}(k+1).
\end{equation*}

Summing together the contributions for all $k$ and normalizing with $2^{n-1}$
(the total number of vectors from which a perturbation may be made, divided
by $2$ to take out symmetric perturbations that have not been counted
separately) gives the sensitivity of $f$ as 
\begin{equation}
s^{f}=n-\frac{1}{2^{n-1}}\sum_{k=0}^{n-1}\left[ (k+1)A_{k+1}+\left( {\binom{n%
}{k}}-A_{k}\right) (n-k)\right] .  \label{sensi_long}
\end{equation}%
Taking into account the fact that $\sum_{k=0}^{n-1}{\binom{n}{k}}%
(n-k)=n\times 2^{n-1},$ this can be simplified to obtain 
\begin{equation*}
s^{f}=\frac{1}{2^{n-1}}\sum_{k=0}^{n-1}\left[ A_{k}(n-k)-(k+1)A_{k+1}\right]
,
\end{equation*}%
which shows the relationship between the coefficients $A_{k}$ and
sensitivity $s^{f}$.

From (\ref{eq:rank}), we can solve 
\begin{equation*}
\left[ A_{k}(n-k)-A_{k+1}(k+1)\right] =p_{n-k}{\binom{n}{k}}(n-k).
\end{equation*}%
Inserting this into (\ref{sensi_long}) results in 
\begin{equation*}
s^{f}=n-\frac{1}{2^{n-1}}\sum_{k=0}^{n-1}\left[ {\binom{n}{k}}(n-k)-p_{n-k}{%
\binom{n}{k}}(n-k)\right]
\end{equation*}%
\begin{equation*}
=\frac{1}{2^{n-1}}\sum_{k=0}^{n-1}{\binom{n}{k}}(n-k)p_{n-k}=\frac{1}{2^{n-1}%
}\sum_{k=1}^{n}{\binom{n}{n-k}}kp_{k},
\end{equation*}%
giving the connection between the rank selection probabilities $p_{i}$ and
sensitivity $s^{f}$.

Interestingly, it has been shown that almost all stack filters are robust in
the sense that all but the central four rank selection probabilities are
nonzero \cite{shmulevich02b}. Moreover, the central two rank selection
probabilities asymptotically dominate over the outer two rank selection
probabilities. Using similar derivations, it is possible to compute the
average sensitivity of a so-called typical monotone Boolean function. The
size of the set of typical monotone Boolean functions asymptotically
approaches the total number of monotone Boolean functions. Equivalently, the
probability that a randomly picked monotone Boolean function is a typical
function approaches unity. Because of the super-exponential growth of the
number of monotone Boolean functions, the asymptotic convergence occurs very
rapidly with excellent accuracy even for as few as seven input variables. It
can be shown that the expected average sensitivity of a typical monotone
Boolean function is given by%
\begin{equation}
\hat{s}^{f}\sim n2^{-n}\binom{n}{n/2-1}\left( 2^{-n/2-1}+1\right) .
\label{eqexpavgsen}
\end{equation}%
for even $n.$The case of odd $n$ is similar to derive, but its expression is
much more cumbersome (see \cite{shmulevich05b}). The expected average
sensitivity of a typical monotone Boolean function is plotted in Figure \ref%
{fig:aveSenBn}. It can be seen that stack filters with as few as five inputs
are already chaotic, in terms of the average sensitivity being interpreted
as a dynamical order parameter.

\begin{figure}[tbp]
\begin{center}
\includegraphics[width=0.8\textwidth]{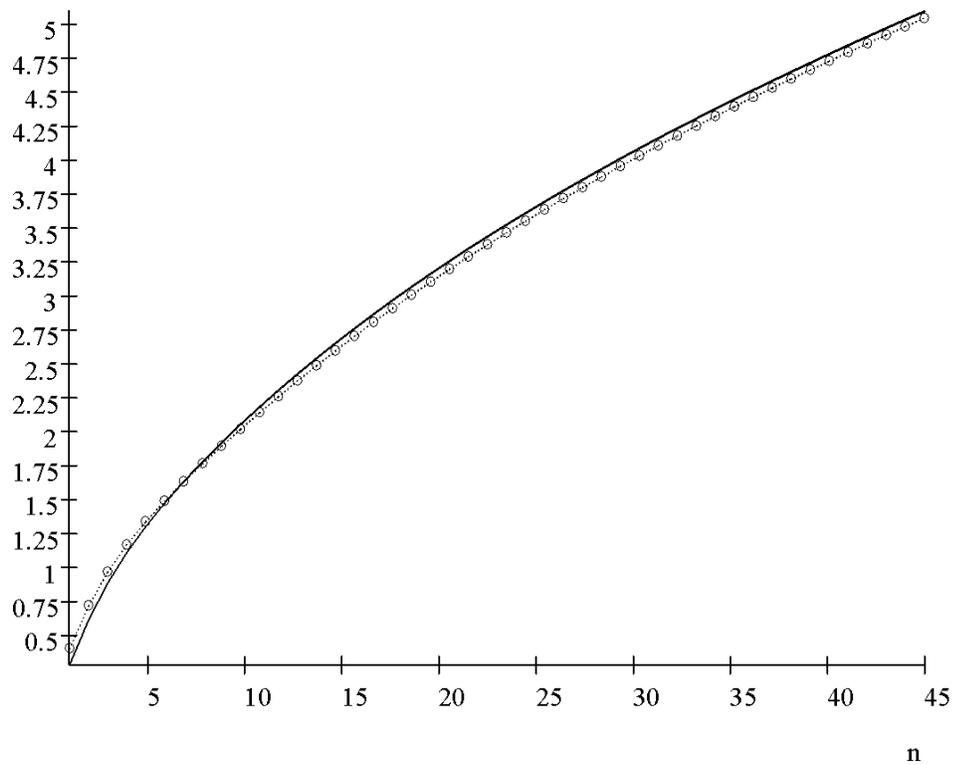}
\end{center}
\caption{The expected average sensitivity of a typical monotone Boolean
function versus the number of variables $n$. The cases of even $n$ (solid
line) and odd $n$ (line with circles) are shown separately.}
\label{fig:aveSenBn}
\end{figure}

\section{Stack filter performance in terms of dynamical behavior}

\begin{figure}[tbp]
\begin{center}
\begin{tabular}{cc}
(a) & (b) \\ 
\includegraphics[width=0.4\textwidth]{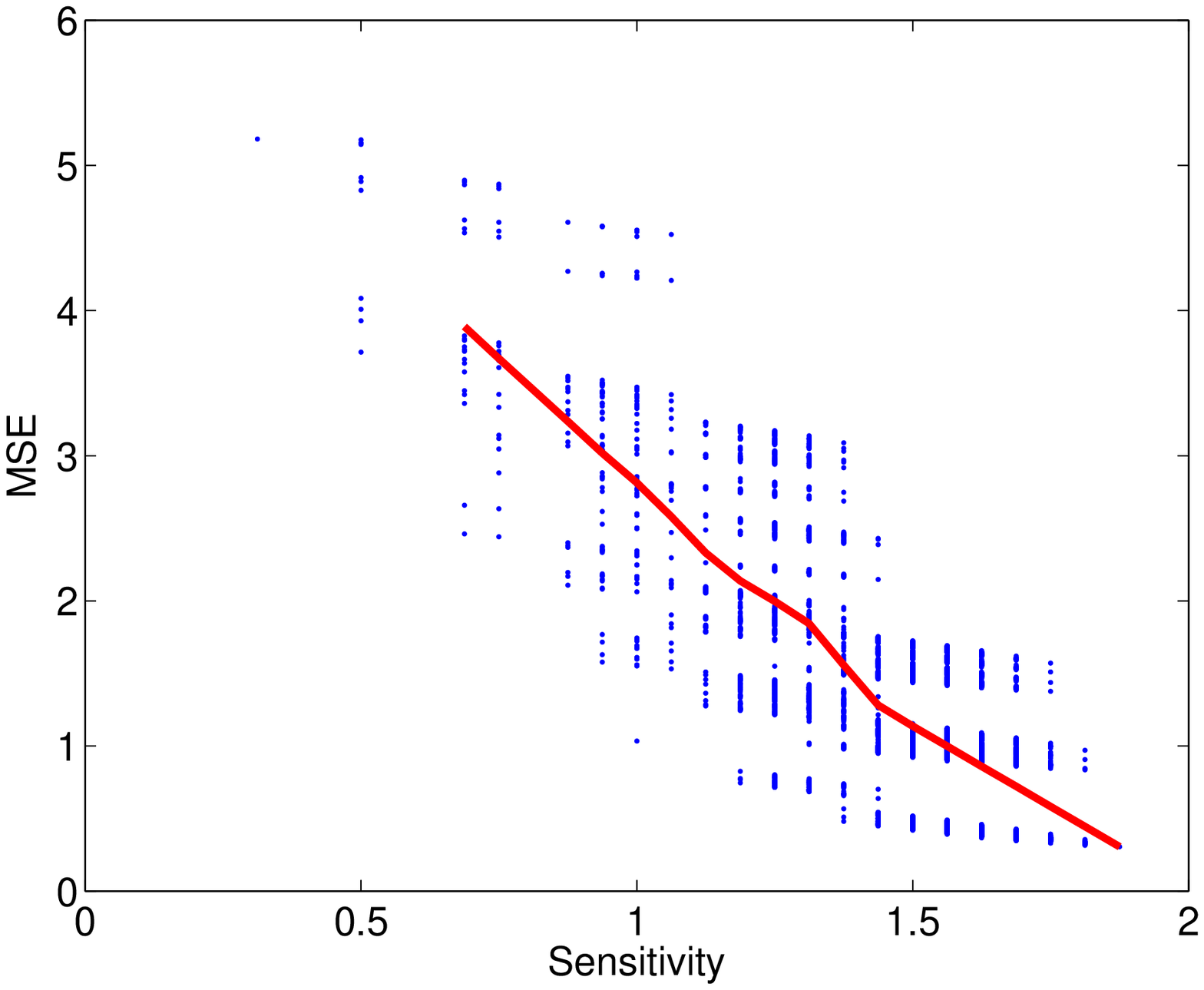} & %
\includegraphics[width=0.4\textwidth]{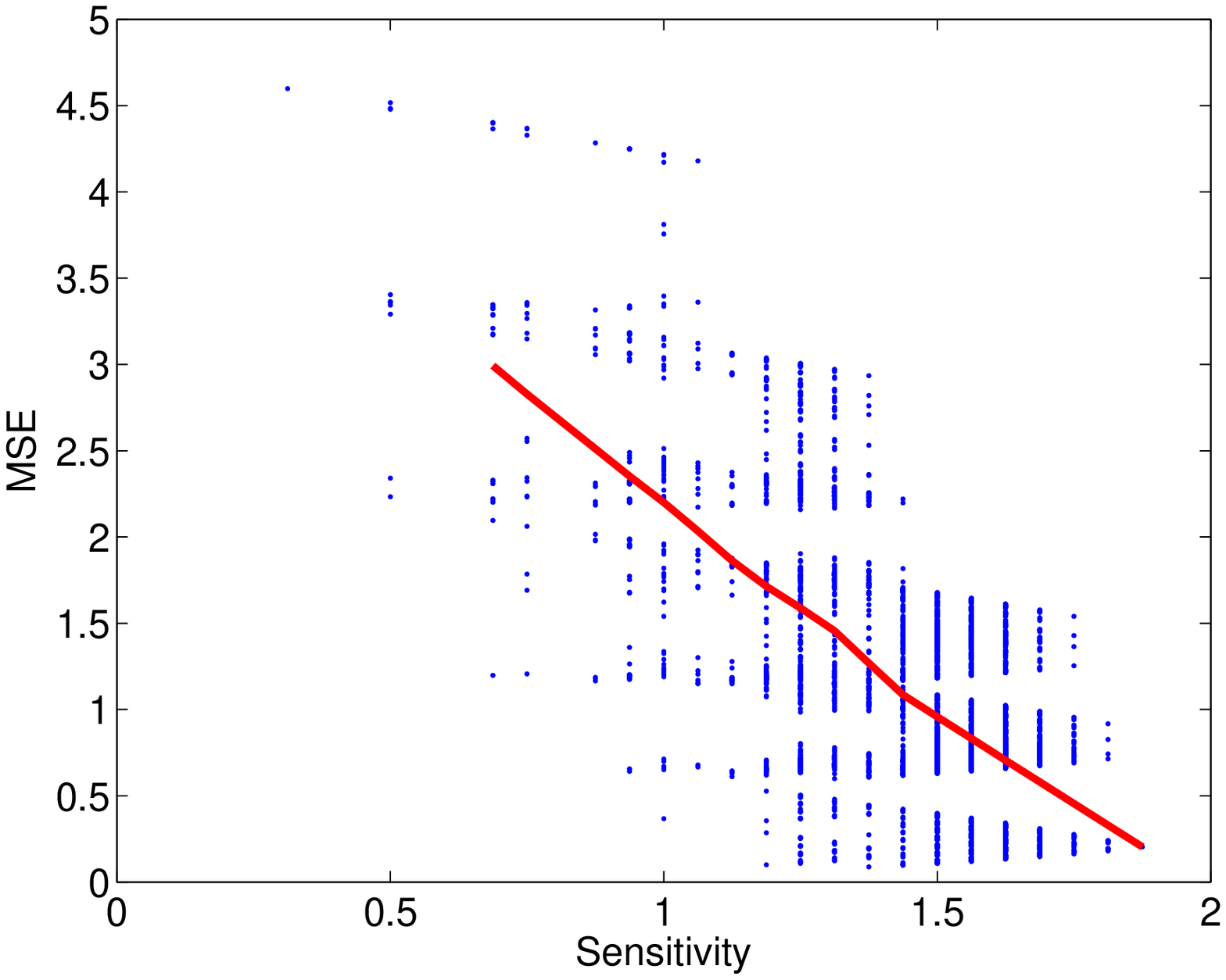}%
\end{tabular}%
\end{center}
\caption{Mean square error versus sensitivity of stack filter. Data is shown
for blocks signal under (a) Gaussian and (b) bimodal noise (see \protect\cite%
{shmulevich00} for details on noise distributions).}
\label{fig:blocks}
\end{figure}

\begin{figure}[tbp]
\begin{center}
\begin{tabular}{cc}
(a) & (b) \\ 
\includegraphics[width=0.4\textwidth]{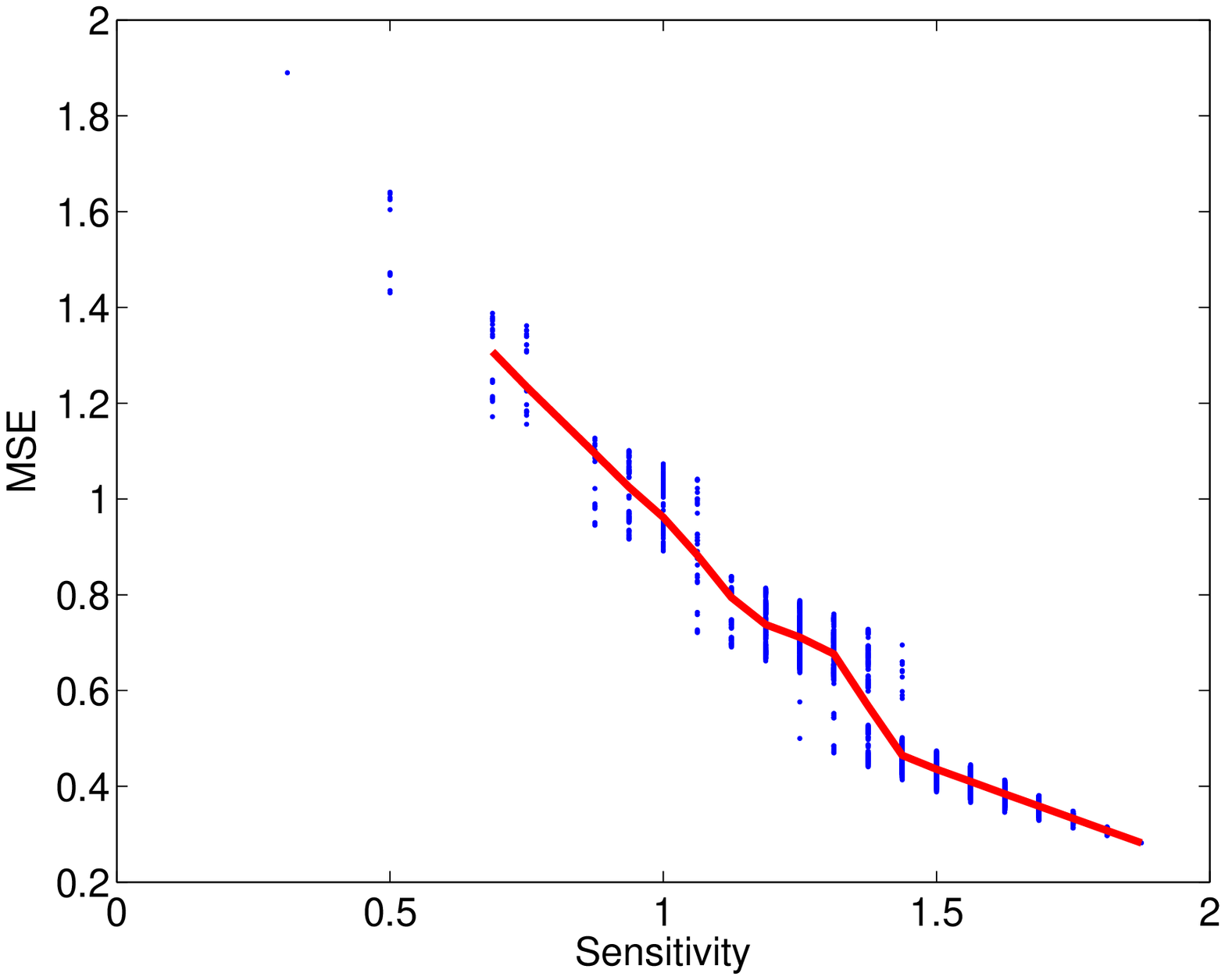} & %
\includegraphics[width=0.4\textwidth]{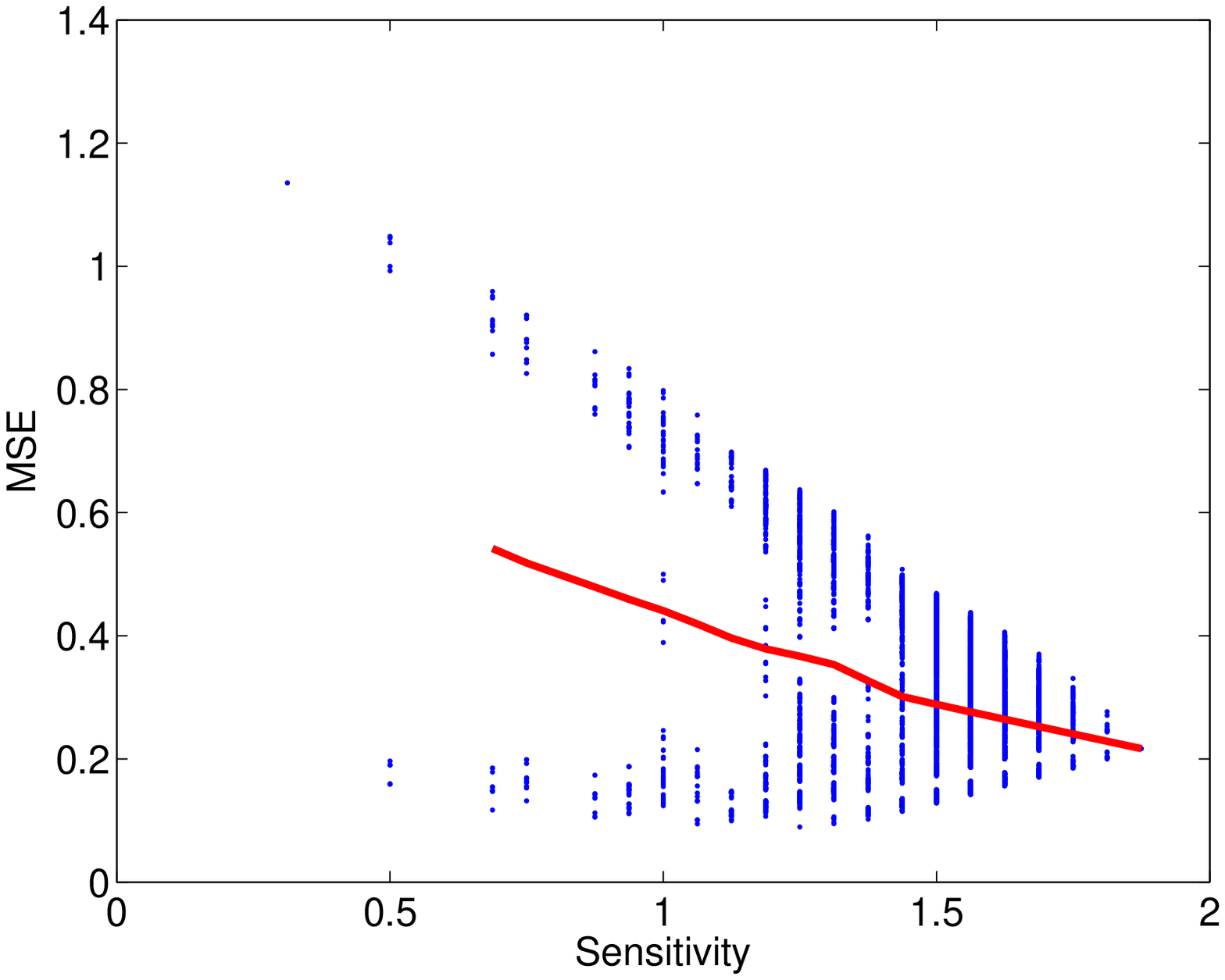}%
\end{tabular}%
\end{center}
\caption{Mean square error versus sensitivity of stack filter. Data is shown
for Heaviside signal under (a) Gaussian and (b) bimodal noise.}
\label{fig:hev}
\end{figure}

\begin{figure}[tbp]
\begin{center}
\begin{tabular}{ccc}
(a) & (b) & (c) \\ 
\includegraphics[width=0.3\textwidth]{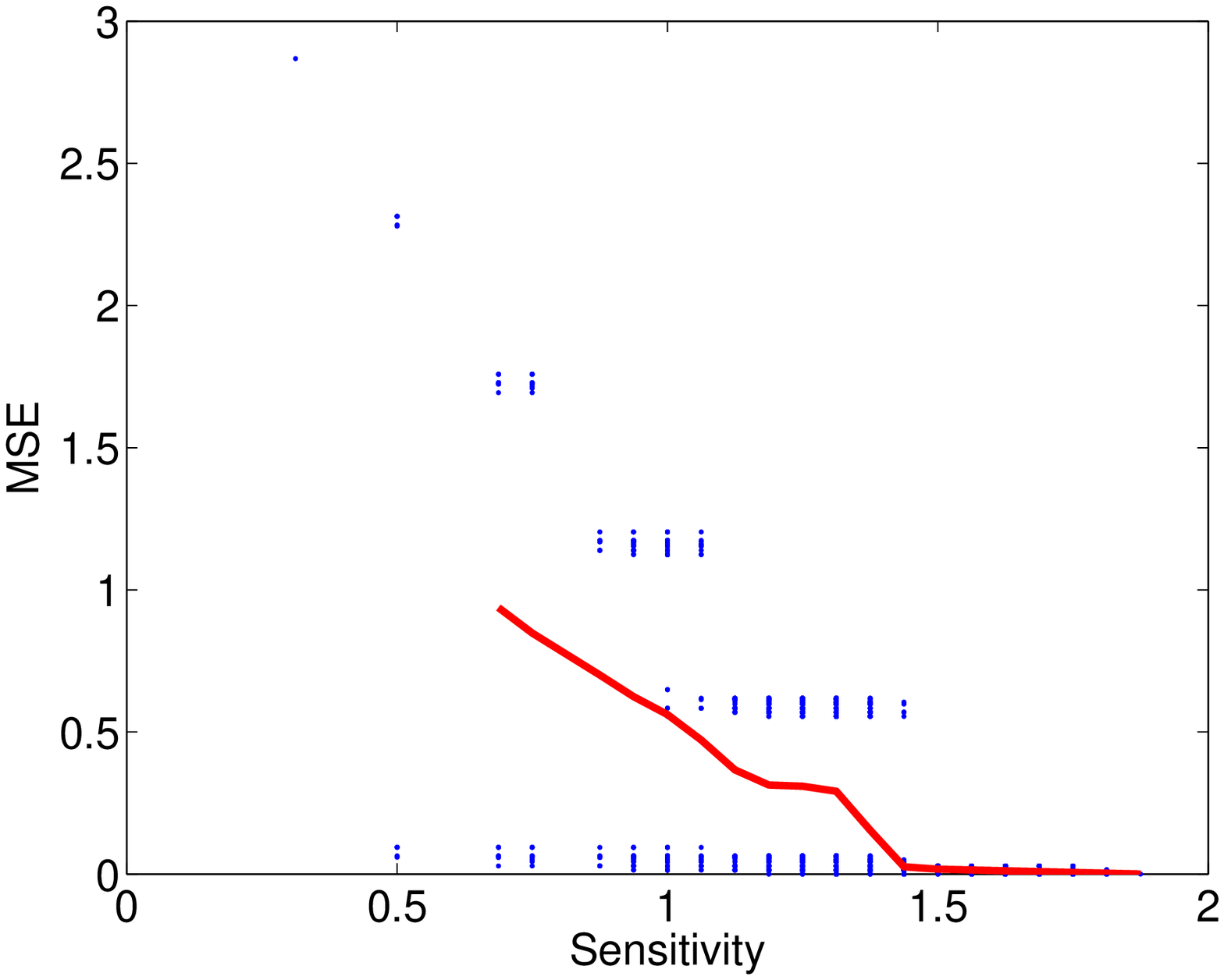} & \includegraphics[width=0.3%
\textwidth]{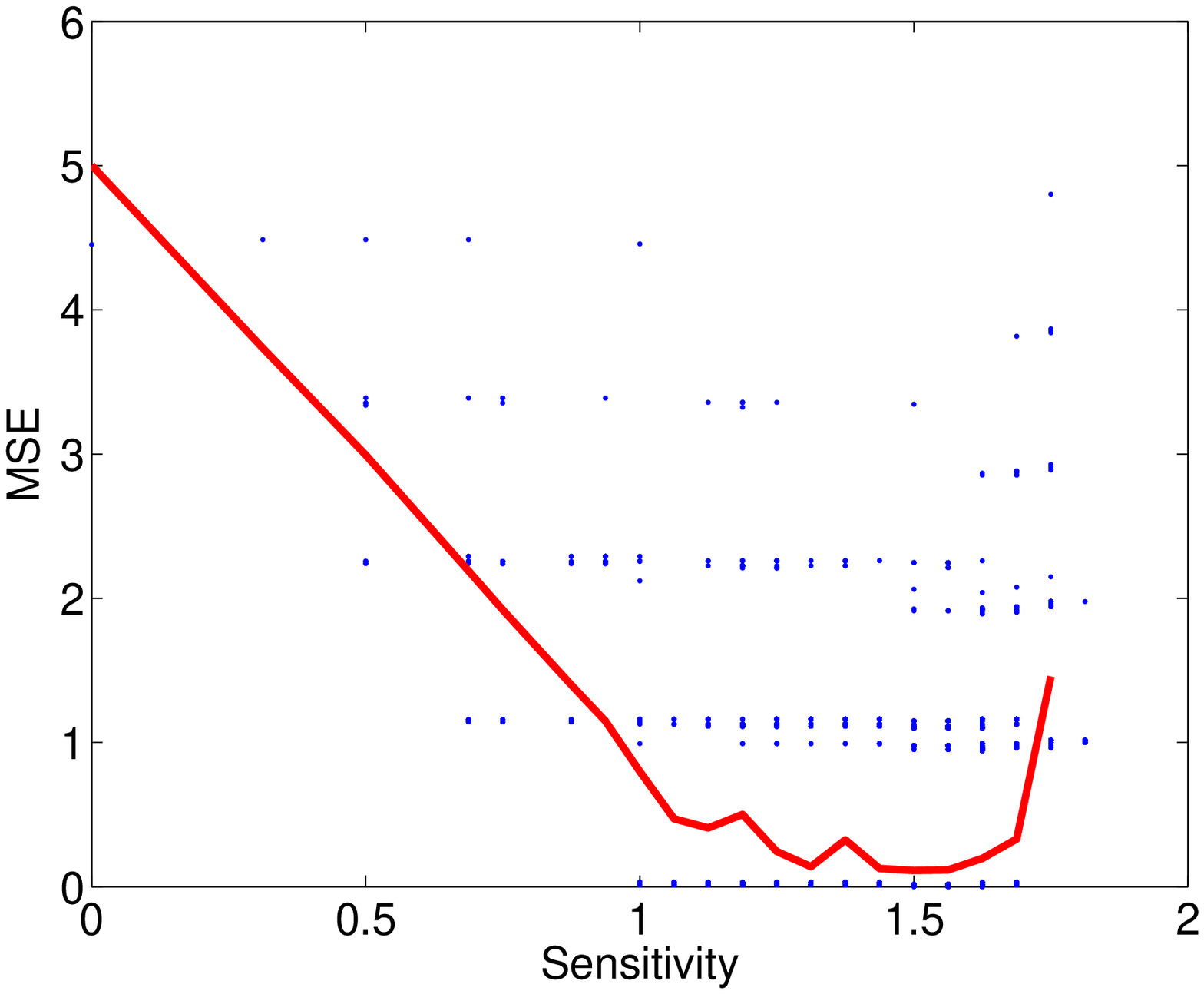} & \includegraphics[width=0.3\textwidth]{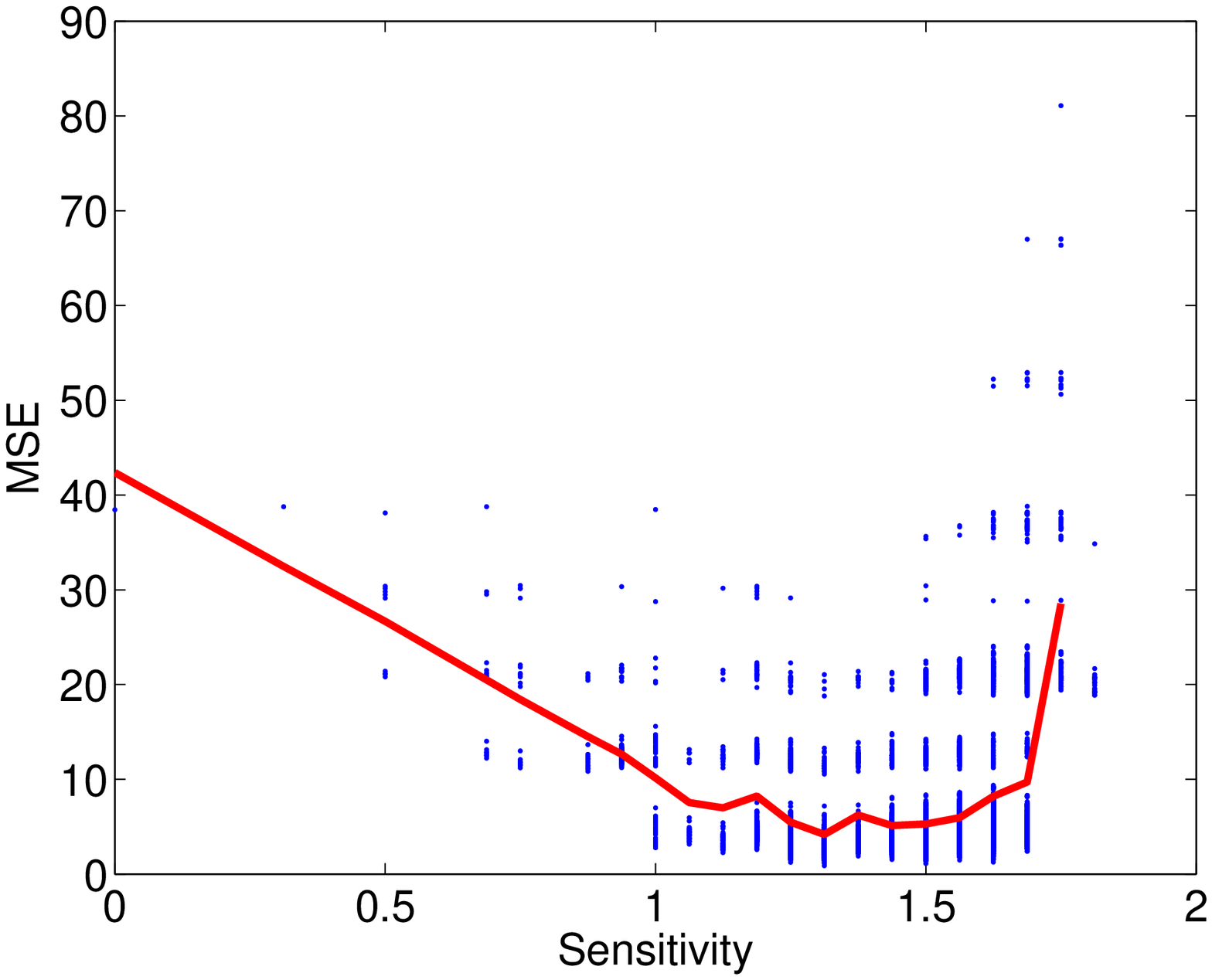}%
\end{tabular}%
\end{center}
\caption{Mean square error versus sensitivity of stack filter. Data is shown
for Heaviside signal under salt and pepper noise with probability (a) 0.001,
(b) 0.01 and (c) 0.1.}
\label{fig:salt}
\end{figure}

To relate the noise suppression properties of a stack filter to the
dynamical behavior, captured by the average sensitivity, we filtered various
test signals and noise distributions with all five-input stack filters.
Figures \ref{fig:blocks} and \ref{fig:hev} show the mean square error (MSE)
as a function of filter sensitivity for blocks and Heaviside signals,
respectively. In addition, Figure \ref{fig:salt} shows the MSE performance
for different degrees of salt and pepper noise.

Based on these results, it is clear that filters that are chaotic, in terms
of their sensitivity, perform best under various noise conditions.
Simulations with salt and pepper noise show that as more noise is added, the
performance of ordered and critical filters deteriorate more that the
chaotic ones.

In light of what we known about the dynamics of random Boolean networks,
this observation is somewhat surprising. A large body of work with Boolean
networks has established that dynamically critical systems are the ones that
strike the optimal compromise between robustness to noise and reliable
propagation of information. Thus, it would be expected that dynamically
critical stack filters would have performed optimally under variable noise
conditions in terms of noise suppression and detail preservation.

There are several aspects of signal denoising that are essentially distinct
from the analysis of dynamical systems. The theory, which relates the
average sensitivity of a Boolean network to the dynamical regimes, is based
on the propagation of small perturbations. That is, to quantify the
dynamical behavior, we measure whether small perturbations that are
introduced into the state of the system are amplified or attenuated, leading
to chaotic or ordered behavior, respectively.

Noise filtering is fundamentally different in terms of what we know about
propagation of small perturbations. A noisy signal is frequently the result
of a large perturbation that has affected the value of majority of the
signal values (consider additive or multiplicative noise that affects every
signal value or pixel). Thus, the insights derived from the small
perturbation analysis of dynamical systems do not necessary hold. The mean
squared error of the filter does not necessary capture the trade-off between
noise suppression and detail preservation that is observed when small
perturbations are studied. Very little is known about the dynamical systems
behavior when only the observations about the response to large
perturbations are available.

\subsection{Stack filter as a generalization of a monotone Boolean network}

An important aspect of stack filters is that they can be viewed as a
generalization of the monotone Boolean network model to a continuos model.
As a stack filter corresponds to a monotone Boolean function and filtering
can be implemented in the form of a Boolean network, the stack filter forms
a network that can be run for continuos signals. This insight gives an
interesting possibility to analyze the relationships between Boolean and
continuos models.

Recently, there have been several attempts to link the properties of Boolean
networks to the behavior of classes of continuous or more detailed discrete
models. For example, the existence of dynamical regimes has been
demonstrated in models other than Boolean networks. The interpretation of
stack filters as generalizations of Boolean networks allows us to study the
connections between Boolean and continuos models directly.

\begin{figure}[tbp]
\begin{center}
\begin{tabular}{ccc}
(a) & (b) & (c) \\ 
\includegraphics[width=0.3\textwidth]{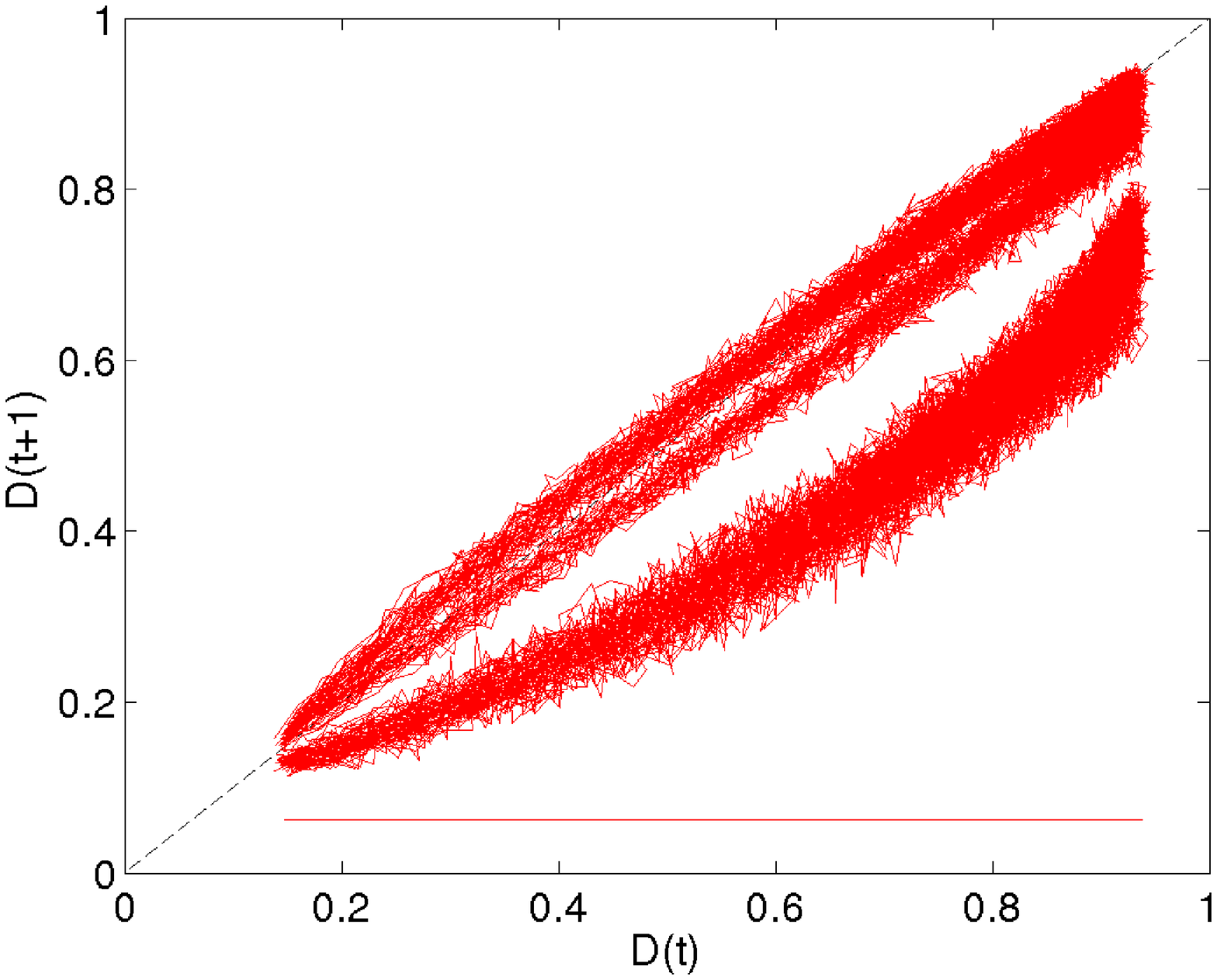} & \includegraphics[width=0.3%
\textwidth]{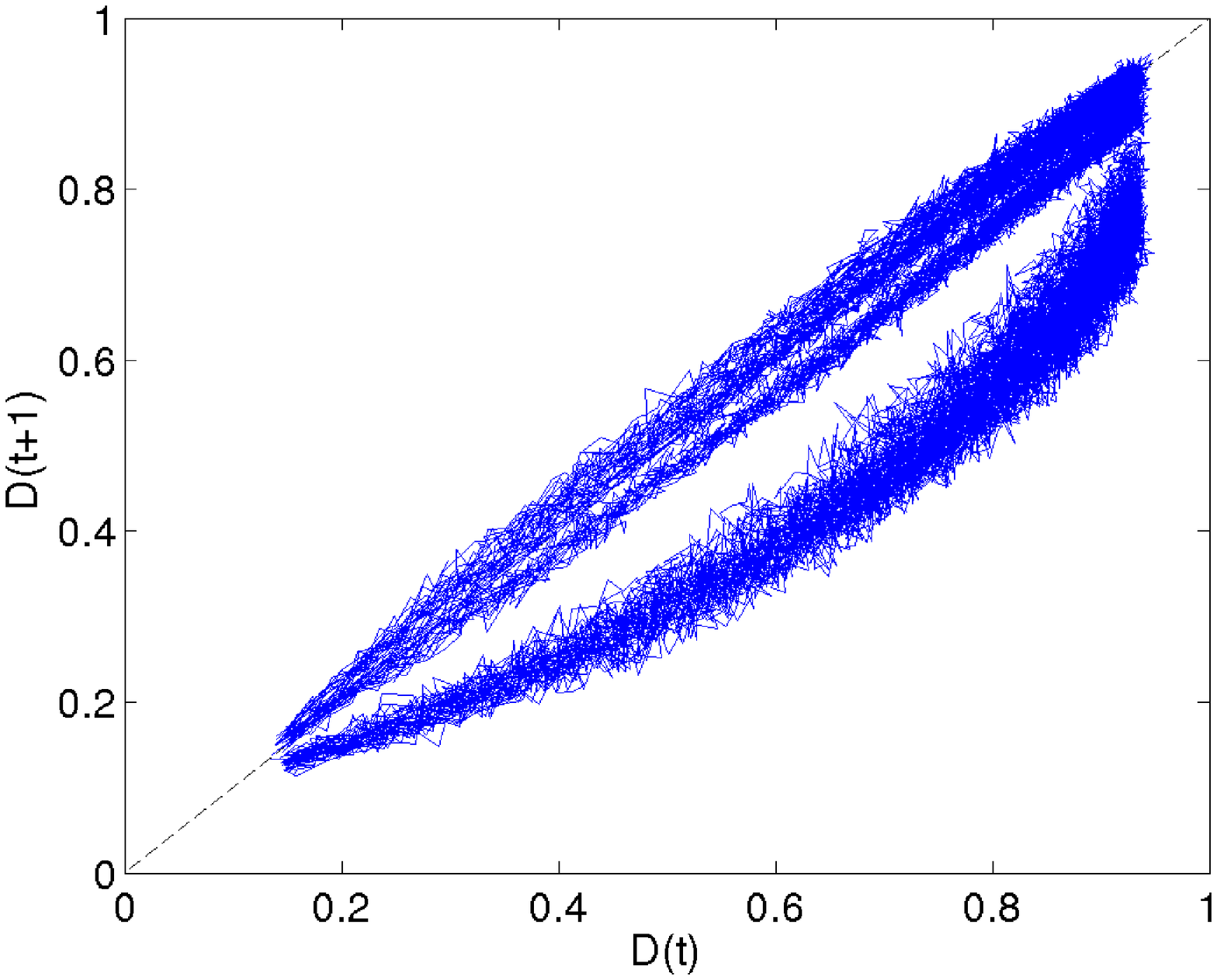} & \includegraphics[width=0.3\textwidth]{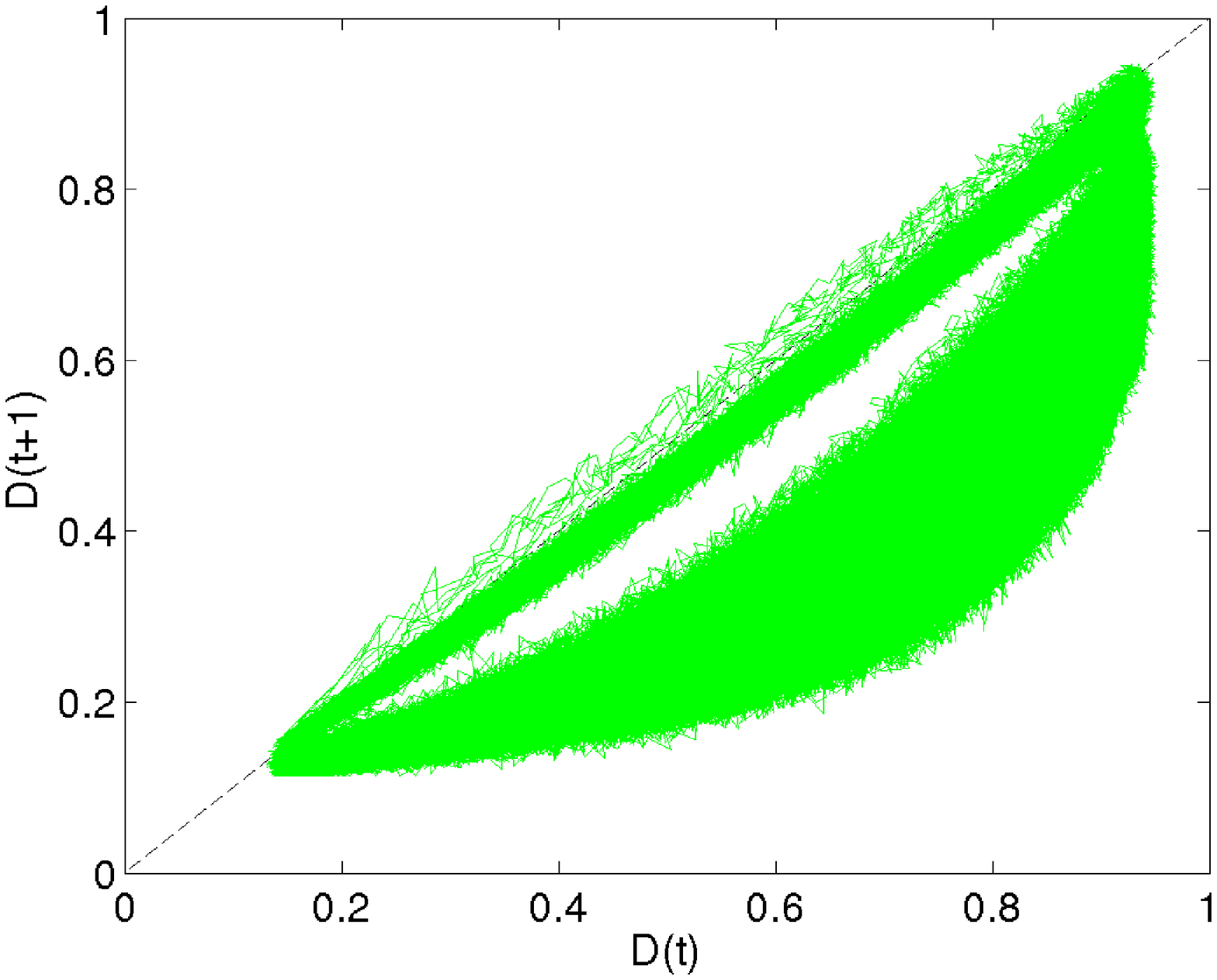}%
\end{tabular}%
\end{center}
\caption{Information theoretical order parameter (Derrida curve) computed
for continuous signals. (a), (b) and (c) show filter classes that are
ordered, critical and chaotic in terms of the average sensitivity of the
Boolean function implementing the stack filter, respectively. All $K=5$
stack filters are shown. $D(t)$ is the information distance between two
input signals and $D(t+1)$ is the distance between filtered signals. Each
line is the averaged behavior of a single stack filter. See \protect\cite%
{nykter08b} for the details on the construction of the Derrida curve.}
\label{fig:ncd}
\end{figure}

In recent work \cite{nykter08b}, we introduced an information theory based
order parameter that can, in principle, be used to quantify the dynamical
behavior of any model class, discrete or continuous. This parameter measures
the average tendency of the dynamical system to attenuate or amplify
informational distances (computed using real world compressors) between
states of the system. We utilize this order parameter to quantify the
dynamical behavior of stack filters in both the binary and continuos forms.
In the binary case, we introduce single bit perturbations to states of the
system with varying probability. In this standard analysis, the information
theory-based order parameter has been shown to accurately capture the
dynamical behavior, characterized by the average sensitivity of the Boolean
function \cite{nykter08b}. For the continuous case, we can introduce noise
in the form of salt and pepper noise with varying probability. Results for
the continuous case are shown in Figure \ref{fig:ncd}.

Remarkably, this analysis suggests that the dynamical regimes of the
networks are different when the input data are changed from binary to
continuous. For example, a Boolean network that is chaotic in terms of
average sensitivity (and confirmed with the information distance -based
order parameter) may not be chaotic in the continuous domain in terms of the
same information distance -based order parameter. This suggests that the
properties that are derived for the monotone Boolean functions do not
necessary hold when the function is generalized to the continuous case.

\section{Conclusions}

We have shown an analytical relationship between stack filter design
coefficients $A_{i}$ and the average sensitivity of the monotone Boolean
function. In addition, we presented a formula for the average sensitivity of
typical monotone Boolean functions. This formula implies that typical stack
filters are dynamically chaotic.

These results suggest a novel perspective that can potentially be utilized
in the design of stack filters for specific filtering tasks. It ties the
design of stack filters to a more general dynamical systems framework. Thus,
we can also utilize stack filter optimization algorithms to design Boolean
networks that are statistically optimal under given noise distributions. It
will be of interest to compare the properties of such statistically optimal
ensembles of networks to the properties of other ensembles under different
noise conditions. By estimating the noise distributions from biological data
and designing optimal Boolean networks may also help us gain insight into
the dynamical behavior of biological systems.

We also studied the behavior of stack filters under various noise
distributions by relating the MSE to the average sensitivity of the filter.
This analysis suggests that chaotic filters perform best under various noise
conditions. This is contradictory to what was expected based on our
knowledge about Boolean networks. However, subsequent analysis shed light on
this observation. Dynamics of Boolean networks are defined under the
assumption of small perturbations while the filtering of noisy signals is a
fundamentally different problem. In addition, our parallel analysis of stack
filter dynamics in Boolean and continuous cases suggests that the dynamical
behavior of monotone Boolean functions does not directly generalize to
continuous systems.

Future work should focus on understanding these aspects in more detail. More
work needs to be done to be able to quantify the dynamics of a system from
large perturbations. It will also be of interest to study the connections
between dynamical behavior of Boolean and continuous models. Stack filters
will be a useful model class for these studies due to the availability of
direct generalizations from the Boolean to the continuous domain.

\bibliographystyle{IEEEtran}
\bibliography{paper}

\end{document}